\documentclass[conference]{IEEEtran}
\usepackage{cite}
\usepackage{amsmath,amssymb,amsfonts}
\usepackage{graphicx}
\usepackage{textcomp}
\usepackage{xcolor}
\usepackage{algpseudocode}
\usepackage{algorithm}
\def\BibTeX{{\rm B\kern-.05em{\sc i\kern-.025em b}\kern-.08em
    T\kern-.1667em\lower.7ex\hbox{E}\kern-.125emX}}
    
\begin{document}

\title{Spectrum Sharing using Deep Reinforcement Learning in Vehicular Networks}

\author{\IEEEauthorblockN{Riya Deshpande}
\IEEEauthorblockA{\textit{School of Computing and Engineering} \\
\textit{University of Huddersfield}\\
Huddersfield, United Kingdom\\
riya.deshpande@hud.ac.uk}
\and
\IEEEauthorblockN{Faheem A. Khan}
\IEEEauthorblockA{\textit{School of Computing and Engineering} \\
\textit{University of Huddersfield}\\
Huddersfield, United Kingdom \\
f.khan@hud.ac.uk}
\and
\IEEEauthorblockN{Qasim Zeeshan Ahmed}
\IEEEauthorblockA{\textit{School of Computing and Engineering} \\
\textit{University of Huddersfield}\\
Huddersfield, United Kingdom \\
q.ahmed@hud.ac.uk}}
\maketitle

\begin{abstract}
As the number of devices getting connected to the vehicular network grows exponentially, addressing the numerous challenges of effectively allocating spectrum in dynamic vehicular environment becomes increasingly difficult. Traditional methods may not suffice to tackle this issue. In vehicular networks safety critical messages are involved and it is important to implement an efficient spectrum allocation paradigm for hassle free communication as well as manage the congestion in the network. To tackle this, a Deep Q Network (DQN) model is proposed as a solution, leveraging its ability to learn optimal strategies over time and make decisions. The paper presents a few results and analyses, demonstrating the efficacy of the DQN model in enhancing spectrum sharing efficiency. Deep Reinforcement Learning methods for sharing spectrum in vehicular networks have shown promising outcomes, demonstrating the system's ability to adjust to dynamic communication environments. Both SARL and MARL models have exhibited successful rates of V2V communication, with the cumulative reward of the RL model reaching its maximum as training progresses.

\end{abstract}


\section{Introduction}
Beyond $5$G encompasses not only wireless communications but also a myriad of applications, including Teleportation, Augmented Reality (AR)/ Virtual Reality (VR), Vehicular Communications, Robotics, and many more~\cite{Ahmed-2020, Ahmed-2020a}. This progressive shift is driving innovations and advancements in diverse fields, ushering in a new era of connectivity and technological possibilities~\cite{Khan-2016, Khan-2016a}. Vehicular communication refers to a network of multiple vehicles communicating with each other or the infrastructure. In other words vehicular communication is also known as Vehicle to Everything (V2X) communication~\cite{va2016millimeter}. V2X network refers to the communication between wireless communication links such as Vehicle-to-infrastructure (V2I) and Vehicle-to-vehicle (V2V). In traditional V2X networks, managing spectrum efficiently poses a significant challenge~\cite{Khan-2016a}. This difficulty arises due to the dynamic nature of the environment, where vehicles are in constant motion and communication necessities fluctuates~\cite{Khan-2021}. The V2X network has limited bandwidth and it is necessary to manage the limited spectrum efficiently. 

Recognizing the complexities involved, the implementation of Artificial Intelligence (AI) has emerged as a promising solution to address these challenges and automate the spectrum management process. One specific AI approach that has shown notable progress in resolving the inefficiencies associated with spectrum sharing in V2X networks is Reinforcement Learning (RL). RL which is a Machine Learning algorithm introduced was aimed to address the challenge of inefficient spectrum sharing, and it has demonstrated significant progress over time. Its ability to learn from interactions and adapt to changing conditions makes it well-suited for optimizing spectrum allocation in real-time scenarios~\cite{Khan-2016, Khan-2016a, Ahmed-2014}.

By leveraging RL algorithms, V2X networks can enhance their adaptability of the dynamic vehicular environment and responsiveness to the requirement of frequency spectrum, ensuring more efficient utilization of available spectrum resources. The interference between channels was gradually reduced after the implementation of the algorithm. This not only addresses the challenges posed by the dynamic movement of vehicles but also contributes to the overall improvement of communication reliability and performance in V2X systems over time. The vehicular networks are based on decentralised decision making where the agents interacts, observes and make decisions on its own unlike other wireless networks where the decisions are taken by the network or infrastructure. In this paper a Deep Reinforcement Learning (DRL) model for the Cellular Vehicle-to-Everything (C-V2X) system is implemented. The system involves vehicles acting as agents that interact with the vehicular environment, generating rewards based on these interactions. 

\section{Related Work}
The progress toward beyond 5G communications is strongly driven by the need to support Vehicular Communication, encouraged by the influence of the 5G Automotive Association (5GAA). Researchers have conducted surveys to examine how the telecommunication industry can contribute to advancing Vehicle-to-Everything (V2X) applications \cite{va2016millimeter}. In a related study \cite{singh2019tutorial}, investigations were undertaken to explore radio communication technologies like millimeter wave (mm-wave), Cellular V2X (C-V2X), and beyond 5G technologies for autonomous vehicles. The study encompassed discussions on challenges associated with these technologies and proposed solutions, specifically addressing the compatibility of wireless communication with vehicular network architecture. Machine Learning (ML) techniques were introduced in the field of vehicular communication. Authors in \cite{tong2019artificial} explored various opportunities and challenges associated with the application of AI and ML techniques in V2X for different applications. Additionally, they provided a description of the vehicular network architecture.

The utilization of RL techniques is expanding over time. The combination of RL with Neural Networks has proven successful in addressing numerous problems. RL is based on the concept of trial-and-error where the agent interacts with the environment and observes the state and receives reward or penalties in terms of feedback. Neural Networks integrated with RL termed as Deep Reinforcement Learning (DRL) has been successful in automating the spectrum sharing process and improved the system efficiency and made the system scalable. A DRL model for spectrum sharing for Device to Device (D2D) communication was implemented by authors in \cite{D2D_Sun} a Multi-Agent Reinforcement Learning (MARL) was designed and simulated.

In V2X communication, there exist multiple heterogeneous networks, and the requirements continually evolve based on the available network size. It is imperative for the system to adapt to these changes and ensure the fulfillment of Quality of Service (QoS) requirements. The authors in \cite{JTian} successfully applied the Multi-agent Deep Reinforcement Learning (MADRL) technique to ensure QoS in vehicular networks, yielding promising results. Additionally, in \cite{Liang2019}, a comparison between two RL models, SARL and MARL, utilizing Neural Networks for spectrum sharing in V2X networks was conducted. The study concluded that RL models achieved a high success probability for V2V transmission. This paper focuses on implementing the RL technique using Neural Network, specifically the DQN model, with a detailed description of the design provided in the next section.

\section{System Model}
Reinforcement Learning (RL) has made significant advancements in the realm of vehicular communication, and various algorithms have been developed for its implementation. This paper focuses on employing a DQN model, which integrates both RL and Deep Learning (DL), to address spectrum sharing challenges in vehicular networks. This section delves into the system's methodology, offering a comprehensive overview.

In the context of RL, the dynamic between the agent and the vehicular environment is mathematically represented by a Markov Decision Process (MDP), as illustrated in Fig 1. The agent observes the current state of the environment, learns from it, and subsequently makes relevant decisions. In simpler terms, the agent acts as the learner, while the environment encompasses all elements with which the agent interacts.\cite{sutton2018reinforcement} The main aim of MDP is to maximize the cumulative reward through the agent and environment interaction by trial-and-error method.

At each discrete time step $t$, the agent is provided with a representation of the environment, indicated as $s_t \in S$, commonly known as the state. Based on this information, the agent takes an action $a_t \in A$, resulting in a numerical reward denoted as $r_{t+1}$, and leading the environment to transition to a new state $s_{t+1}$. The Markov property asserts that the current state encapsulates all relevant information about both previous and future states, with future states being independent of past states. In the realm of Reinforcement Learning (RL), the primary objective for an agent is to maximize future rewards, even though a high reward at the present state does not necessarily guarantee superior rewards in the future.
\begin{figure}[!t]
     \centering
    \includegraphics[width=0.9\linewidth]{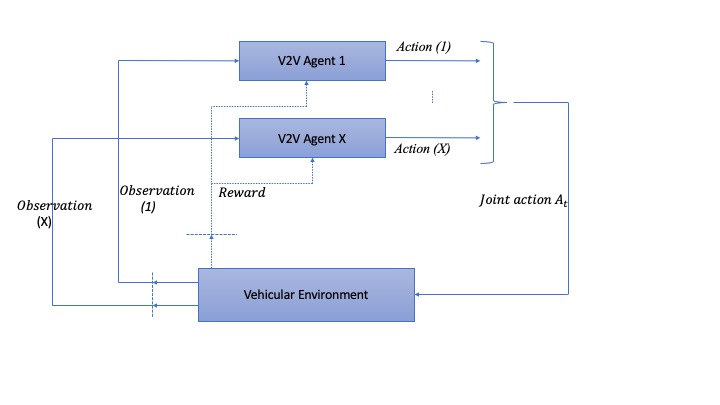}
    \vspace{-0.8cm}
    \caption{Agent Interaction}   \label{fig:enter-label}
 \end{figure}
In this paper, we consider a spectrum sharing scenario illustrated in Fig. 2 involving $Y$ V2I links and $X$ V2V links, where $Y=\{1,2,..,y\}$ and $X=\{1,2,..,x\}$. The V2I links are high frequency communication links and are connected directly to the base station for data services such as live streaming or surfing internet. It is assumed that V2I links Y are preoccupied and reused by V2V links X.

The channel power gain for x V2V links over y V2I link is denoted in (1).

\begin{equation}
    g_x[y] = a_xl_x[y]
\end{equation}
The variable $l_y$ signifies the power of small-scale fading, influenced by the frequency, while $a_y$ denotes the value of large-scale fading, which remains independent of the frequency. The channel models selected for implementation are detailed in \cite{ii2007winner}.
\begin{figure}[!b]
    \centering
   \includegraphics[width=0.9\linewidth]{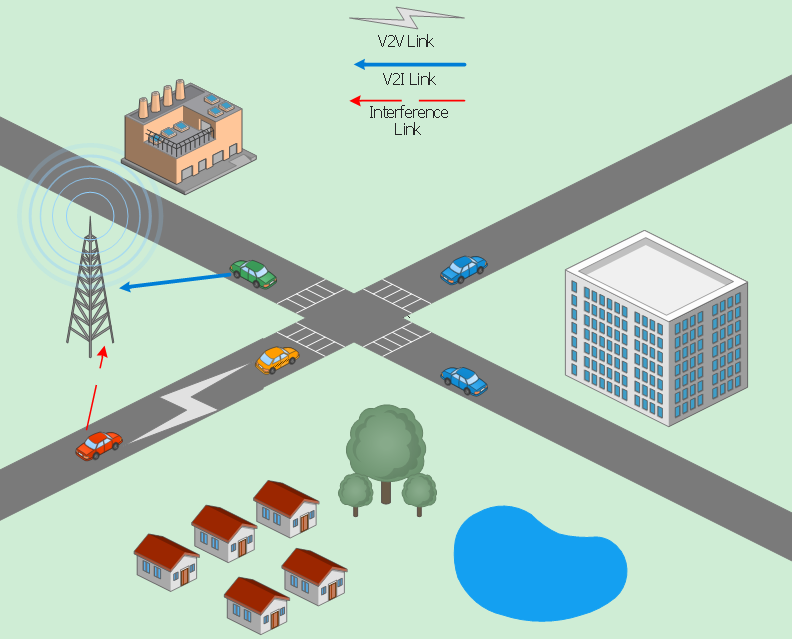}
    \caption{Illustration of system model}
    \label{fig:enter-label}
\end{figure}

\subsection{Multi-Agent Reinforcement Learning for Spectrum sharing}

RL encompasses two noteworthy models: Single Agent Reinforcement Learning (SARL) and Multi-Agent Reinforcement Learning (MARL). SARL addresses scenarios where a single agent interacts with the environment, generating rewards. Conversely, MARL involves multiple agents interacting simultaneously with the environment. The primary objective of these models is to enhance spectrum efficiency and ensure scalability, allowing a maximum number of devices to communicate efficiently with minimized interference. In these complex and uncertain situations RL algorithms have provided promising results. Within the spectrum sharing context, numerous V2V links endeavor to access the constrained V2I spectrum that is already in use. In this scenario, V2V links function as agents interacting with the vehicular environment, making observations that are subsequently utilized in policy design. The V2V agents refine strategies for spectrum sharing and power control based on their experiences derived from the current state of the environment. The MARL model, as applied in this paper, is segregated into training and testing phases, with a detailed description provided for the fundamental processes involved in MARL-based spectrum sharing.

\subsubsection{State and Observation}
In RL, the agent explores the vehicular environment for discrete time t, and make some observations $Z_t^{(x)}$ is received by agent of current state $S_t$. The state space is mathematically described in MDP. Here observation $Z_t^{(x)} = O(S_t,x)$ for individual agent x. The Observation space  includes own channel information, information of local channel, and interference links from all transmitters in the network.

\subsubsection{Action}
The process of spectrum sharing involves the selection of a spectrum band and the control of transmission power. In the simulation, the power of X V2V links has been restricted to four levels, namely $[23, 10, 5, -100]$ dBm. Specifically, a power level of -100 dBm signifies zero transmit power. The actions in this context are represented by pairs of spectrum and power.

\subsubsection{Reward Design}
The primary objective of the model is to optimize the success probability of V2V payload transmission and enhance the V2I transmission capacity. The RL algorithm aims to maximize the cumulative reward, consequently increasing the likelihood of achieving maximum data transfer for V2V links. The calculation of the cumulative reward is carried out during the training episodes 

\subsubsection{Training}
The training process is segmented into 3000 episodes, each with a V2V payload generation time of $T=100ms$ and varying sizes $P=[1,2,...]*1060$ bytes. During each episode, the agent observes the environment and, based on these observations, selects suitable actions along with an exploration rate. The DQN for each V2V agent comprises hidden fully connected layers with 500, 250, and 120 neurons, respectively. The neural network illustrated in Fig 3 is employed to learn the policy, mapping states to actions and generating state-action pairs. A learning rate of 0.001 is utilized, and the mini-batches are divided into 2. The algorithm for MARL was implemented for training of the agents. The training and testing models were simulated separately. Relevant actions are selected after the observation process and the outcomes are stored into replay memory buffer. The channel fading is updated to reduce the train loss and interference.

\begin{figure}[!t]
    \centering
   \includegraphics[width=0.9\linewidth]{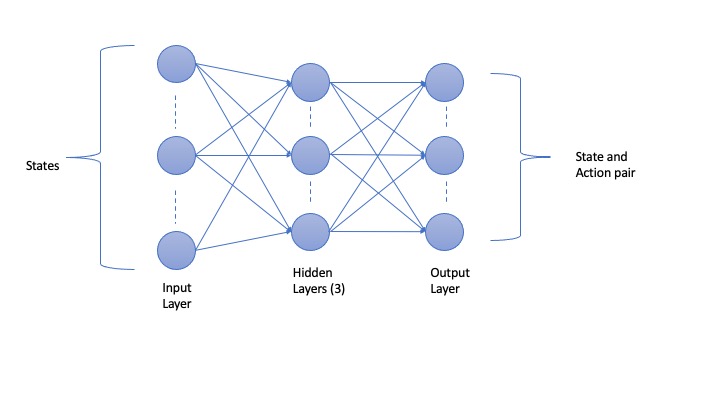}
   \vspace{-1cm}
    \caption{Deep Q Network}
    \label{fig:enter-label}
\end{figure}

\begin{algorithm}[htb!]
\caption{Algorithm for MARL \cite{LiangSPAWC}}\label{alg:cap}
\begin{algorithmic}
\State Start environment simulator, generate agent (vehicle) links
\State Set up neural network for the agents individually
    \For{\text{each episode}}
        \State Update location of vehicles
        \State Update $\alpha$ (large scale fading)
        \State Reset $P_x = P$ \& $T_x = T$, where $x \epsilon X$
        \For{\text{time step t}} 
            \For{\text{agent x}} 
                \State Observe the state $Z_t^{(x)}$
                \State Choose an action based on $\epsilon$-greedy policy
            \EndFor
            \State  Each agent takes individual actions and receives 
            \State feedback in the form of reward $R_{t+1}$
            \State Update the small-scale fading in the channel
            \For{\text{agent x}} 
                \State Observe the new state $Z_{t+1}^{(x)}$
                \State Store the outcome in the Replay Memory $C_x$
            \EndFor
        \EndFor
        \For{\text{agent x}}
            \State mini-batches from $C_x$ 
            \State Optimize error using stochastic gradients
        \EndFor
    \EndFor   
\end{algorithmic}
\end{algorithm}

\section{Results and Discussion}
In this section results for the model implemented are discussed, these results are generated in python using TensorFlow and NumPy frameworks. The channel model for V2I and V2V links is described in 3GPP documentation as well as the cellular scenario used to implement the environment simulator is mentioned in \cite{network20113rd}. 
\begin{figure}[htb!]
    \centering
    \centerline{\includegraphics[width=0.9\linewidth]{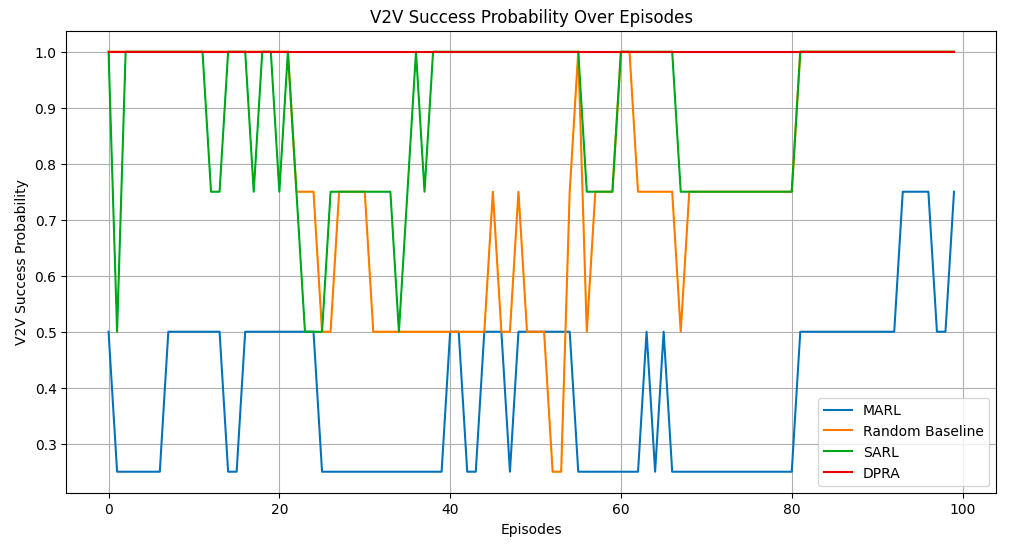}}
    \caption{V2V success probability}
    \label{fig:enter-label}
\end{figure}
Fig 4 illustrates the V2V success probability across test episodes, offering insights into the system's performance. The plot indicates the likelihood of establishing a successful V2V link. While the SARL model exhibits the highest probability, the MARL model demonstrates a more consistent and stable payload transmission.
\begin{figure}[htb!]
    \centering
    \centerline{\includegraphics[width=0.9\linewidth]{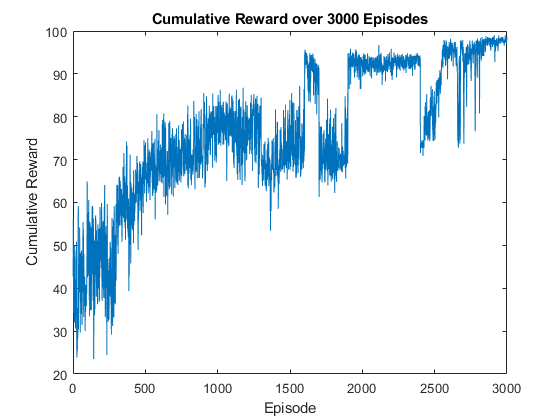}}
    \caption{Cummulative Reward}
    \label{fig:enter-label}
\end{figure}
The cumulative reward, as shown in Fig 5, is computed with a constant payload size of $P=2,120$ Bytes during training the model. The channel model specifications are outlined in \cite{ii2007winner}. In ideal scenario it is assumed that RL has to maximize the cumulative reward so that maximum data is transferred. The cumulative reward is calculated while training the agents and it is the sum of all rewards received in one episode. The ROC converges after 2500 episodes. Fluctuations in the plot represents that results are influenced by channel fading. The path loss and channel fading parameters are mentioned in \cite{ii2007winner}.

\begin{table}[htb!]
    \centering
    \caption{Results}
    \label{tab:my_label}
    \begin{tabular}{|c|c|c|}
        \hline
        Model & Average V2I  & Success Probability \\
        &Transmission Rate&of V2V\\
        \hline
        MARL & $30.55264$ & $0.4225$\\
        \hline
        SARL & $40.01552$ & $0.9975$\\
        \hline
        Random & $37.04313$ & $0.9075$\\
        \hline
    \end{tabular}
  \end{table}

The findings presented in Table 1 showcase the Average V2I transmission rate, measured in Mega bits per second (Mbps), and the V2V success probability. The conclusion drawn is that the RL model exhibits a higher probability of successfully establishing V2V links. The results indicate that the DQN model adeptly adjusts to the dynamic and uncertain vehicular environment. This adaptability contributes to the overall efficiency of the system, as successful communication links are established across the network. The transmission rate is notably influenced by the optimization of decisions made by DQN agents.

\section{Conclusion}
The work was focused on the implementation of spectrum sharing model in vehicular networks using deep reinforcement learning algorithm. The study involves research on RL for Vehicular Network and the implementation of a DQN model. The results shows the increase in successful transmission of V2V links and the cummulative reward is maximized while the training is progressed. It is concluded that the average transmission rate for RL model is comparatively higher than other algorithm.

\bibliographystyle{IEEEtran}
\bibliography{V2X}

\end{document}